# Vibration motions studied by Heterodyne Holography

**F. Joud[1], F. Verpillat[1], P.A. Taillard[2], M. Atlan[3], N. Verrier[3,4] and M. Gross[4]**
[1]*Laboratoire Kastler Brossel - UMR 8553 CNRS-UPMC-ENS 24 rue Lhomond 75005 Paris France*
[2]*Conservatoire de musique neuchâtelois, Avenue Léopold-Robert 34; CH-2300 La Chaux-de-Fonds; Switzerland*
[3]*Institut Langevin UMR 7587CNRS ESPCI ParisTech, 1, rue Jussieu, 75005 Paris France*
[4]*Laboratoire Charles Coulomb - UMR 5221 CNRS-UM2 place Eugène Bataillon 34095 Montpellier France*
*gross@lkb.ens.fr*

**Abstract:** Playing with amplitude, phase and frequency of both reference and signal arms, heterodyne holography is well adapted to vibration analysis. Vibration sidebands can be imaged and stroboscopic measurement sensitive to mechanical phase can be made.
**OCIS codes:** (090.1760) Computer holography; (200.4880) Optomechanics; (040.2840) Heterodyne; (100.2000) Digital image processing

There is a big demand for full field vibration measurements, in particular in industry. Different holographic techniques have been used to analyze vibrations, the most simple and most common one being time averaged holography [1]. We have developed heterodyne holography [2] that is a variant of phase shifting holography [3] that control the reference arm phase with acousto optic modulators. This technique can be used to study vibration very efficiently.

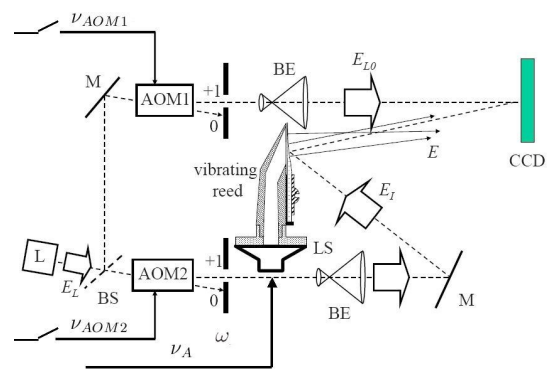

**Fig. 1** – Typical vibration heterodyne holography setup. L : main laser ; AOM1, AOM2 : acousto-optic modulators ; M : mirror ; BS : beam splitter ; BE : beam expander ; CCD : CCD camera ; LS : loudspeaker exciting a clarinet reed through the bore of a clarinet mouthpiece at frequency $\nu_A$.

Figure 1 shows a typical heterodyne holography setup applied to vibration analysis. The modulators (AOM1 and AOM2) make possible to control electronically both the amplitudes, phases and frequencies of the fields of both the reference ($E_{LO}$) and signal arms ($E_I$ and $E$). Because of the vibration motion at frequency $\nu_A$, the signal field $E$ can be developed into carrier $n = 0$ and sideband $n \ne 0$ field components $E_n$ of frequency $\nu_n = \nu_0 + n\nu_A$ where $\nu_0$ is the frequency of the illumination optical field $E_I$:

$$E = \sum_{n=-\infty}^{+\infty} E_n e^{2\pi v_n t} \quad \text{with} \quad E_n = j^n J_n(A) e^{2\pi v_n t}$$

$J_n$ is the Bessel function of order $n$, $A$ is the phase modulation amplitude and $j^2 = -1$. By a proper choice of the AOMs frequencies ($v_{AOM1}$, $v_{AOM2} \sim 80 MHz$) it is then possible to detect selectively each sideband $n$. For example, to perform the 4 phases detection of sideband $n$, one must make : $v_{AOM1} - v_{AOM2} = v_A + v_{CCD}/4$ where $v_{CCD}$ is the CCD camera frame frequency.

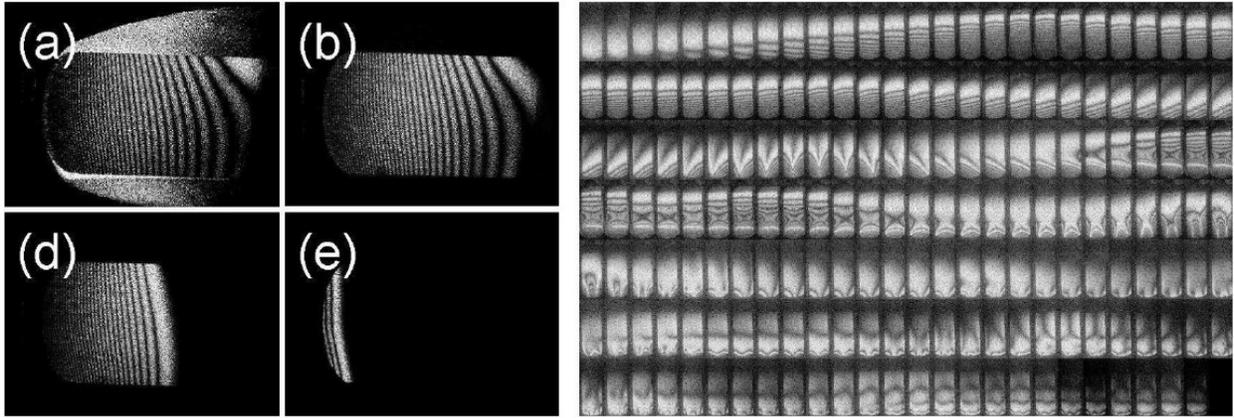

**Fig. 2** – (Left hand side) Reconstructed holographic image of a clarinet reed vibrating at frequency $v_A = 2143$ Hz. Carrier image (a) with $n = 0$, and sideband images (b,c,d) with $n = 1$ (b), $n = 20$ (c) and $n = 100$ (d). (Right hand side) $n = 1$ holographic images of a clarinet reed for $v_A = 1.4$ to 20 kHz by steps of 25 cents (181 images ordered from left to right, continued on the next row).

Figure 2 (left) shows typical vibration images that have been obtained at fixed frequency $v_A = 2143$ Hz for sideband rank $n$ varying from $n = 0$ to $n = 100$ [4]. The time averaged fringes obtained for $n = 0$ are shifted toward the tip of the reed when increasing the sideband rank $n$. Since all frequencies ($v_{AOM1,2}$ and $v_A$) are driven by numerical synthesizer, one can sweep the vibration frequency and perform an automatic acquisition of the vibration field maps as shown on Fig.2 (right) [5].

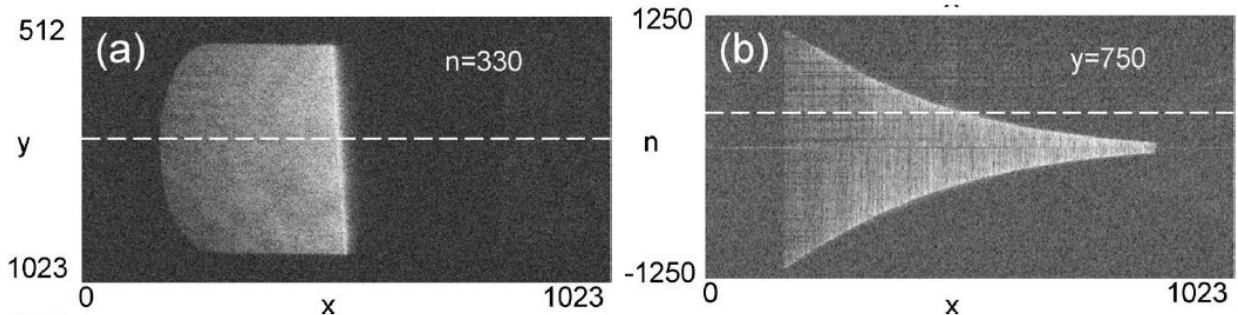

**Fig. 3** – (a) $x, y$ image reconstructed with sideband $n = 330$, with a large amplitude of vibration. (b) $x, n$ images corresponding to cuts of 3D data along the horizontal white dashed line of (a).

By recording a cube of data of coordinate $x, y$ and $n$, and by performing cut along the $x$ direction, it is also possible to analyze large amplitude vibration that cannot be studied by time averaged holography, because the time averaged fringes are too tight . Figure 3 show an clarinet reed example with an amplitude of vibration of $\pm 60 \mu m$ on tip of the reed [6].

It is also possible to use the AOMs to turn on and off the signal and reference arms at the vibration frequency, and to perform by the way stroboscopic holography. In the case of large amplitude vibration, one can get a images the vibration velocities as shown on Fig.4 (left) [7]. For small amplitude, one can get image of the vibration motion by slowly drifting the phase of the stroboscopic illumination and by performing a double demodulation on the data as shown on Fig.4 (right) [8]. Note that we have not consider here the case of very small amplitude vibration. In that case, heterodyne holography is also extremely useful since, as mentioned by Psota et al. [9], the detection of the sideband of harmonic rank $n = 1$ is intrinsically well adapted to the detection of small vibration amplitude $A$. For $n=1$, the holographic signal varies like $J_1(A) \cong A/2$, while, for time averaged holography ($n=0$), the signal varies like $J_0(A) \cong 1 - A^2/4$.

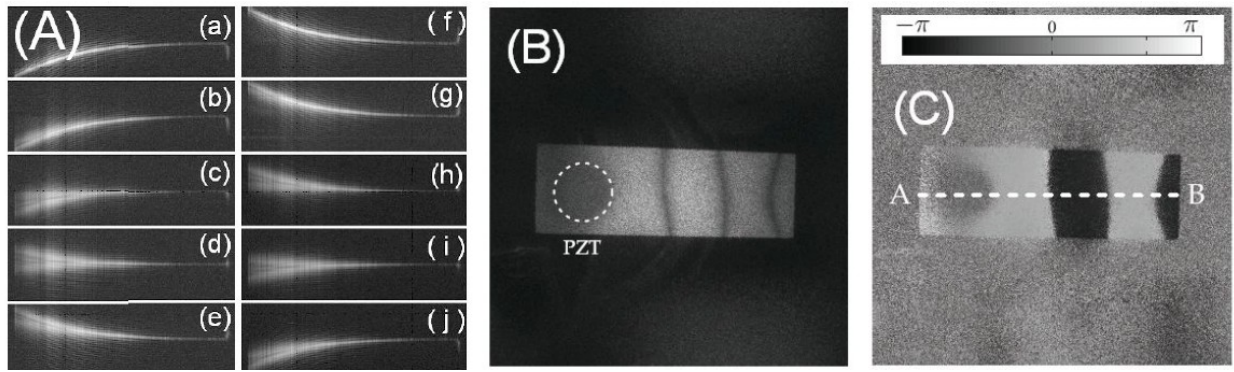

**Fig. 4** – (A) $x, n$ cuts of 3D data (analog to the Fig.3(b) cut) made for successive strobe time. (B,C) Amplitude (B) (black : nodes, white : anti nodes) and phase (C) image of a vibrating sheet of paper with drifting strobe illumination and double demodulation.